\documentclass[pre,floatfix]{revtex4-1} 
\usepackage[pdfpagemode=UseNone,pdfstartview=FitH,colorlinks=true]{hyperref}
\usepackage{enumitem}
\usepackage{url}
\setlistdepth{9}
\usepackage{graphicx}
\usepackage{epstopdf}

\renewcommand{\[}{\begin{equation}}
\renewcommand{\]}{\end{equation}}

\begin{document}
\title{Computational mechanisms in genetic regulation by RNA}

\author{J. M. Deutsch}
\email{josh@ucsc.edu}
\affiliation{Department of Physics, University of California, Santa Cruz CA 95064}

\begin{abstract}
The evolution of the genome has led to very sophisticated and complex regulation.
Because of the abundance of non-coding RNA (ncRNA) in the cell, different species will promiscuously associate with each
other, suggesting collective dynamics similar to artificial neural networks.
Here we present a simple mechanism allowing ncRNA to perform computations equivalent
to neural network algorithms such as Boltzmann machines and the Hopfield model. 
The quantities analogous to the neural couplings are
the equilibrium constants between different RNA species. The relatively rapid equilibration
of RNA binding and unbinding is regulated by a slower process that degrades and creates new RNA.
The model requires that the creation rate for each species be an increasing function of
the ratio of total to unbound RNA. Similar mechanisms have already been found to exist
experimentally for ncRNA regulation.  With the overall concentration of RNA
regulated, equilibrium constants can be chosen to store many different patterns, or many different
input-output relations. The network is also quite insensitive to random mutations in equilibrium
constants. Therefore one expects that this kind of mechanism will have a much higher mutation rate
than ones typically regarded as being under evolutionary constraint.
\end{abstract}
\maketitle
 
\section{Introduction}

The overwhelming majority of transcripts in the human genome  produce non-coding RNA (ncRNA) and these
have been under intensive investigation in recent 
years~\cite{ENCODEpilot,Kapranov2007,MercerDingerMattick,ENCODE,djebali2012landscape,lee2012epigenetic} 
which has revealed many functions. However, research to date has still only scratched the surface of the mechanisms
involved with these transcripts.

Aside from specific mechanisms, it is useful to take a step back and ask at an algorithmic level,
what all of this extra RNA might be capable of doing, given the constraint that the mechanisms be biologically plausible.
The author proposed~\cite{Deutsch14} that a general way of understanding
many of ncRNA's functions was to have these molecules act {\em collectively}.
By collective behavior, what is meant is that the actions of any one piece
of the genomic circuitry is influenced by a large number of different molecules.
This contrasts with the usual way of understanding biological regulation, where
specific molecules will interfere, suppress, or promote, gene expression. This is
most often how elements in {\em cis}-regulation are described. With
collective mechanisms, such specific pathways cannot explain function. The 
system needs to be considered in its entirety for the correct genomic behavior to emerge.

The ``connectionist" model of human cognition and machine learning, has been considered
in rather early work in the context of many biological processes including gene 
regulation~\cite{ mjolsness1991connectionist}.
One can model regulatory elements physically, where binding and unbinding
are controlled by equilibrium constants. The binding of regulatory proteins in such networks is
generally thought to be quite specific, and these models are more akin to circuit
diagrams with a few connections in and out of every element. The program pursued here is
to understand if it is biologically tenable to instead have a large number of molecules
present that have  much less specific
interactions, and yet can function in a precise way, regulating many of the myriad functions
that take place in the cell. Of course, this is not meant to suggest that all functions operate
this way, but that this collective mechanism could also be operating. 

This paper describes a surprisingly simple mechanism for achieving
this kind of collective regulation, where perhaps thousands of RNA species 
bind to each other promiscuously, yet this results, or
indeed is responsible, for a high level of computational complexity.
This is motivated by the developments in artificial intelligence
that have come about from consideration of similar collective
models~\cite{HertzKroghPalmer}. One of the most general kinds of
models in this class is the ``Boltzmann machine"~\cite{BoltzmannMachine},
which in a certain limit, described below, becomes the so-called
``Hopfield
model"~\cite{LittleNeural,LITTLE1975115,LITTLE1978281,hopfield1982neural,amit1985spin,amit1987information}.

Earlier work by the author~\cite{deutsch2016associative} described a way of mapping the Hopfield model onto
promiscuously binding RNA molecules. However it was not clear how such a mapping could be made compatible with 
the biological and physical requirements.
The work here broadens the class of physical systems, and also shows how the
mechanism can be greatly simplified to make it much more credible that such a mechanism
would be able to evolve. Other recent work~\cite{poole2017chemical} proposes more direct methods
for constructing chemical analogies of Boltzmann machines, but so far it is not clear how this
is related to biology.

To make the proposal here biologically plausible, its mechanisms should involve functions
similar to those already known to exist. There are two mechanisms necessary in what follows in order for it to
work. The first is that different chemical species bind and unbind in accord with statistical mechanics.
The second is the existence of molecular mechanisms to selectively transcribe species of RNA depending
on the fraction of it that is bound to other species. There are many ways of achieving the right
mathematical form and
there are many forms for this dependence that will work. This sort of behavior is fairly 
typical in many biochemical subsystems. 
A speculative proposal for accomplishing this would be that this process takes place
with little genomic involvement.  There is machinery capable of replicating RNA, 
similar to RNA-dependent RNA polymerase (RdRP),  which is essential for the viability of many
viruses~\cite{spiegelman1965synthesis}, but also appears to exist in humans~\cite{kapranov2010new}.
Furthermore the transcription rate of RdRP should depend on the relative amount of bound to unbound
polymer for every molecular species. A more conventional approach uses the effect that ncRNA has on 
genomic transcription factors. This general kind of mechanism has been
observed~\cite{takemata2017role} in different situations. These possibilities are discussed in Sec. \ref{sec:MechForCreation}.

One of the main points of this work is to illustrate that there may
be very different principles lurking in biological systems of
which we are currently unaware. These would not be apparent to us
from the sophisticated arsenal of experimental techniques we now
use to understand genetic regulation. These tools are primarily
designed to tease out specific interactions of a few components
from the myriad that exist. On the other hand, hypothetically, to observe collective regulation,
one needs to be able to examine thousands of components simultaneously,
each one having a minuscule effect, but collectively, they produce
precisely controlled regulation. An analogy with artificial neural
circuitry might make this point clearer. In pattern recognition
systems, where one desires to classify different images, most of
the neurons fire in response to essentially all images that are
presented. A single neuron is involved in the recognition of hundreds
of thousands of images. Yet by precisely controlled collective
interactions between units, very specific and accurate classification
is achieved.  Even in the case where all neurons can be probed
simultaneously, it can be very difficult to understand how the
circuitry operates, because the collective interaction of many
components is not conducive to the kinds of explanations used in
more normal digital circuitry. The same is expected for biological
neural circuits as well.

\section{Relation to Boltzmann Machines}
\label{Sec:RelToBoltzMach}

\begin{figure}[htb]
\begin{center}
\includegraphics[width=0.5\hsize]{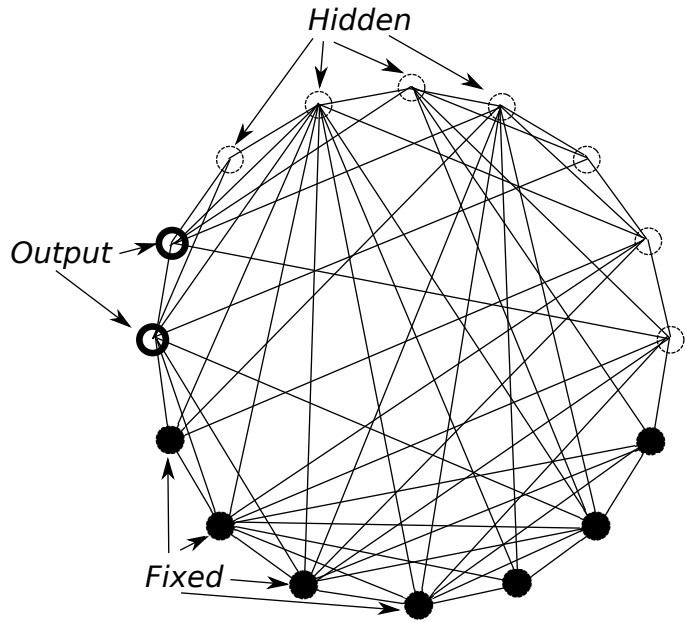}
\caption
{ 
Illustration of a Boltzmann machine network. The black circles are inputs to the network, that are
fixed in value during processing. The unfilled circles on the left are output units (in this case
there are two), whose final
values depend on the state of the input neurons. The remaining unfilled circles are hidden units
that are used to process the input units.
}
\label{fig:BM}
\end{center}
\end{figure}

The purpose of a Boltzmann machine~\cite{BoltzmannMachine,hinton1983,hinton_optimal} 
is to learn a set of input/output pairs, and generalize from that information. If a set of inputs
is presented, a corresponding set of outputs should be retrieved. This is accomplished as follows.

Consider a set of variables, often referred to as ``spins", $s_i$ that can take on only the values $\pm 1$,
but can change their values over time, as the system is updated, for example using the
Metropolis Monte Carlo algorithm at some finite temperature.

They interact via a connectivity matrix  $J_{ij}$ that couples spins $i$ and $j$. The couplings
are chosen by a method outlined below, to optimally give the correct outputs.
To do this updating, one writes down an energy function, or Hamiltonian, for this system
\[
\label{eq:SpinH}
H = -\sum_{i=1,j=1}^N J_{ij} s_i s_j
\]
Out of these $N$ spins, there $N_f$ spins that can sometimes be fixed. This is sometimes referred to as the ``clamping" phase. 
These $N_f$ spins are often referred to as ``visible" units. 
When they are fixed, they have constant values 
during the updating procedure. 
These represent the training set of data for the machine.
One can think of their effect as externally imposing constraints
on the dynamics of the other $N_v \equiv N-N_f$ variable spins. Thus the first $N_v$ spins $s_1,\dots,s_{N_v}$ can vary
and the last $N_f$ spins $s_{N_v+1},\dots,s_N$ are, sometimes, frozen. Of the last $N_f$ spins, we can regard
$N_i$ as inputs and $N_o = N_f-n_i$, as outputs. These are the input/output pairs mentioned above.
This is depicted in Fig. \ref{fig:BM}, where the black circles represent the input units (that is
spins), and the outputs are the black unfilled circles. The dashed unfilled circles represent the
hidden units. The interactions $J_{ij}$ are depicted by straight lines connecting the spins.

The algorithm starts by fixing these $N_f$ spins to one of the
input/output pairs while the system goes through many updating cycles. Then the system
is allowed to run with the $N_f$ visible spins now being {\em unclamped} so that they are no longer fixed and are
updated in the same way as the rest of the spins. Comparing the statistics of the unclamped and clamped simulation
allows one to evolve the weights $J_{ij}$, moving the system closer towards the optimal set of couplings. 
As this happens, the temperature of the system is also slowly decreased.
This is an application of simulated annealing~\cite{kirkpatrick1983optimization}.

The next step in this process is to change the fixed spins to another input/output pair and
the above unclamping and annealing process repeats. Eventually, the connectivity matrix will have evolved to one that results
in a system that has optimally learned these input/output pairs. That is, if one of the input sets is presented, it responds
by giving the corresponding output.

The Hopfield model~\cite{LittleNeural,LITTLE1975115,LITTLE1978281,hopfield1982neural,amit1985spin,amit1987information}
considers the case of no hidden units. In that case, there are $M$ input/output pairs to be learned. There is no logical
distinction here between inputs and outputs, and any subset of the spins could be presented, with the
expectation that the rest would correctly flip to the desired outputs. If we wish to learn $M$ separate 
spin configurations $t_1^\alpha, \dots,t_N^\alpha$, $\alpha = 1,\dots,M$, then a Hebbian rule can be used
to explicitly write down the couplings without going through any simulated annealing procedure,
\begin{equation}
\label{eq:Hebb}
J_{ij} = {\cal N} \sum_\alpha t_i^\alpha t_j^\alpha 
\end{equation}
with a normalization factor $\cal N$, that varies depending on the author, and will be chosen
later on in what follows.
The number of patterns that can be reliably stored is proportional to
$N$~\cite{amit1985spin} but this will also depend on the correlations between the patterns.
The mean field solution of these equations at inverse temperature $\beta$  are
\begin{equation}
\label{eq:Hopfield}
s_i = \tanh(\frac{\beta}{N}\sum_j J_{ij} s_j) .
\end{equation}
With the Hebbian couplings of Eq. \ref{eq:Hebb}, and
statistically $t_i^\alpha$'s, the choice
$s_i = t_i^\alpha$ can be shown~\cite{HertzKroghPalmer} to satisfy these equations.

One can also use the same couplings in a Boltzmann machine with hidden units. This will not
be optimized over the choice of spin values for the hidden spins, but for this choice of coupling,
it will lead to a recall of all the visible units.

\section{The System}

RNA is used in an enormous number of ways in biology. The majority of transcripts in human cells do
not directly code for proteins, but are non-coding~\cite{ENCODE} with functions that are mostly not well understood. 
Here we make the conjecture that that much of this RNA is involved in the kind of collective regulation
described above. 

The system studied here is a collection of $N$ RNA species that interact through base pair binding and unbinding.
Each molecule can bind to itself and other molecules. RNA-RNA interactions in physiological
conditions allow for the formation of secondary structure, and therefore must also allow for the
binding to different molecules, as such interactions are of identical strength and mathematical form. The amount 
of non-coding RNA (ncRNA)
in a cell are quite substantial~\cite{cabili2011integrative}.
We therefore expect that these RNA molecules are strongly interacting, and that furthermore that
they will bind strongly to other molecules such as some proteins. This is discussed further in
section \ref{sec:MagRNAint}.

The inputs to the cell, such as signaling molecules, are well known to affect the transcription of DNA
to RNA, and in particular, messenger RNA (mRNA). 
Because of the strong interaction between RNA molecules, this in turn will affect the concentrations of all
of the RNA. Some of these other RNA molecules will be involved with protein translation. 
By promoting or suppressing protein translation, these RNA concentrations will affect the function of the cell. 
Thus the cell inputs are ``processed" by a complex system of DNA, proteins, and RNA, to produce or modify cell outputs.
What we will investigate here, is the possibility that it is the strong interactions between
the RNA molecules that underly the complex computations used by the cell in determining how it will respond to 
different inputs. 

The basic linking between inputs and outputs, is similar to the standard regulatory mechanisms involving
binding of regulatory proteins along different sites on the genome~\cite{buchler2003schemes}. 
In contrast with the RNA proposal above, 
such bindings require much more specificity, in much the same way as a conventional
digital circuit requires precise connections between its elements. A system of promiscuously binding
RNA molecules in all likelihood, is incapable of such specificity, and instead must resort to collective
behavior, in analogy with artificial neural networks.

To achieve such collective computation, we use two basic components.
The first is that
there is a chemical equilibrium between $N$ molecular species undergoing reversible reactions. 
Such an equilibrium is well understood~\cite{Reif}. 
We assume that because of the relatively high concentration of molecules, the system can quickly reach equilibrium
concentrations. The steady state concentrations of different species is what we will be interested in studying here.

On the other hand, the lifetime of RNA molecules is finite, and this means that in steady state, they require
creation. 
The way that this happens is crucial to the second component to this model and is discussed in
detail below. 
This is a subtle problem and certainly requires experimental verification. In order
for our set of RNA molecules to perform sophisticated computations, we looked for a simple rule that
would be biochemically plausible. What we will find is that the creation rate of a species depends on the ratio of bound to
unbound RNA. We will discuss possible ways this can be achieved later in Sec. \ref{sec:MechForCreation}.

In the following sections, we will show that these two rather simple assumptions involving RNA equilibration
and creation, can perform collective
computations. We first will consider an intermediate model where the mathematical relationship with learning algorithms
is the most apparent. We will then show how this can be simplified further to come up with a more biologically
plausible mechanism.

\subsection{Chemical  equilibration}

We consider $N$ different chemical species of RNA that bind and unbind at rates that depend on their
base pair sequences. For example, complementary sequences will be most strongly bound.
For the moment, assume that there are fixed total concentrations of each species $C_1, C_2, \dots, C_N$.
We will denote the corresponding unbound concentrations as $\rho_1, \rho_2,\dots,\rho_N$. 
We will assume binary reactions between molecules $i$ and $j$
\begin{equation}
i + j \rightleftharpoons ij
\end{equation}
and that there are no higher order reactions present. Including more complex reactions should not preclude the scenario
presented here from working, and is an interesting topic for further investigation.
We will denote the concentration of two molecules $i$ and $j$, that are bound together, by $\rho_{ij}$.

The equilibrium constant~\cite{Reif} for such a reaction is $K_{ij} = \rho_{ij}/\rho_i\rho_j$.
This implies~\cite{Deutsch14}
\begin{equation}
\label{eq:ConcentrationModification}
\rho_i =\frac{C_i}{1+\sum_j\rho_j K_{ij}}
\end{equation}
for $i=1,\dots,N$.

In this model, the set of equilibrium constants $K_{ij}$, is fixed during the lifetime of a cell, as
would be expected. It is posited that these evolve through
mutation, to be able to take on arbitrary values, within some physical limits. Because binding
between two different molecules will take place preferentially along certain species specific regions, 
there are enough degrees of freedom to these binding affinities to be chosen independently.
Even if there is some dependence, there are many possible choices for couplings that lead to
useful learning.

\subsection{Creation of new RNA}
\label{sec:CreationOfNewRNA}

As mentioned above, degraded RNA molecules must be replaced by new ones, and the most
subtle part of the mechanism proposed here is how molecule production is regulated.
In this model we have two requirements related to RNA creation. First, 
that the rate will be controlled by the fraction of total to unbound molecules
of the same species. Second, we require a mechanism to regulate the total concentration of
RNA molecules. This latter requirement is fairly uncontroversial, and this can be regarded as a
homeostatic feedback mechanism.
More specifically, the first requires that the generation of a particular RNA species increases
as the ratio of total to unbound RNA increases. We will discuss how these mechanisms could operate
in more detail in Sec. \ref{sec:MechForCreation} and will briefly summarize the two main
scenarios that could give rise to this kind of regulation.

As mentioned in the introduction, RNA-dependent RNA polymerase (RdRP), can directly
copy RNA molecules without the presence of DNA. For a given species, we require that the rate of copying will
depend inversely on the amount of free RNA present. 

A more likely explanation for how such a dependence could be achieved is through DNA cis-regulatory
elements, regulating RNA transcription. There are already similar kinds of regulation that have been
observed~\cite{takemata2017role}, and this will be discussed further in Sec. \ref{sec:MechForCreation}.  However this
kind of regulation is of course not evidence for the existence of the kind of computation proposed here,
but suggests that this kind of mechanism would be worthwhile to look for experimentally.

In the first model we will consider, the transcription rate will depend on other factors as well,
(see Sec. \ref{sec:CreationRate}), but we will show later in Sec. \ref{sec:MoreUnivReg},
that we can dispense with these additional dependencies and produce a
model with a transcription rate as described above, making this much more biologically plausible.

The point of this model is as a proof-of-concept. In addition, there are many variants, some of which
have already been mentioned, such as higher order interactions, that could also perform collective
computation. The main point of the following analysis is to make the case that this sort
of mechanism is plausible.

\subsection{Dynamics of concentration and transcription rates}

Assuming no degradation or creation of RNA, the system will go to equilibrium
concentrations given by Eq. \ref{eq:ConcentrationModification}, which determines the unbound 
RNA concentrations given the total concentrations of all the RNA molecules.
However because of degradation and creation, the actual concentrations will differ
from the equilibrium case.

In the framework described here, certain RNA species, for example mRNA, will act as inputs and we will assume that
these have fixed values, while the remaining species will time variation in their bound and unbound concentrations,
due to RNA degradation, creation, and the interactions with other molecules.
Out of the $N$ species of RNA, we can say the $N_v$ of the concentrations can vary and $N_f$
of them are are fixed, with $N_v+N_f = N$.

If a closed system is not initially in equilibrium,
the unbound concentrations will vary in time, asymptotically approaching the equilibrium values.
The actual dynamics will be very complex and there will be a spectrum of
relaxation times associated with the RNA concentrations' dynamics. 

We assume that binding and unbinding takes place on a timescale much shorter
than the lifetime of an RNA molecule, and that this 
allows us to describe dynamics with only one relaxation time $\tau_\rho$ through a standard
first order kinetic equation.
\[
\label{eq:FirstOrderRho}
\tau_\rho\frac{d\rho_i}{dt} = -\rho_i + \frac{C_i}{1+\sum_j\rho_j K_{ij}}
\]
for $i=1,\dots,N$.  Note that for a general linear system, we expect a spectrum 
of relaxation times. All of these times are assumed to be very
short compared to the other processes described below, and therefore such details
on a longer timescale are unimportant. This will be justified further in Sec. \ref{sec:MagRNAint}.

We now quantify the mechanism of RNA creation discussed above. We 
assume a degradation timescale $\tau_C$ that is independent of molecular species. 
The rate of transcription of the {\em ith} species is regulated by a process that depends on both the
total concentration $C_i$ of a species, and all of the unbound $\rho$'s.  
\begin{equation}
\label{eq:FirstOrderC}
\tau_C \frac{d C_i}{dt} = - C_i + f(C_i,\{\rho_k\}) 
\end{equation}
for $i=1,\dots,N_v$. 
Later we will show how this dependence can be considerably simplified to make it
more biologically plausible.
We expect that the process of degradation and production takes place at a much
slower time scale than the molecular equilibration mentioned above, so that $\tau_C \gg \tau_\rho$.
This is born out by estimates using empirical data, as discussed in Sec. \ref{sec:MagRNAint}.
The function $f$ gives the rate at which molecules of type $i$ are being
created through the kind of mechanism described above in Sec. \ref{sec:CreationOfNewRNA} and in
Sec. \ref{sec:MechForCreation}.

The remaining $C_i$, $i=N_v+1,\dots,N$, will act as inputs as described above, and those
concentrations will not vary in time. Processes external to the ones that we considering, are maintaining those levels,
for example, the transcription of mRNA molecules that are acting as fixed inputs.
%We will also consider variants of this where the corresponding $\rho_i$'s are also held fixed. But
%maintaining the total concentration of a species appears, at least superficially, to be biologically 
%the most plausible scenario.

\subsection{Relation to Boltzmann Machines}
\label{sec:RelTiBoltzMach}

We will now connect the machine learning system discussed in Sec. \ref{Sec:RelToBoltzMach}
to the genomic system above. The variables of interest for the Boltzmann Machine are
the spin variables $s_1,\dots,s_N$. We relate these to the concentration of unbound RNA $\rho_1,\dots,\rho_N$.
In this case the $s_i$ will no longer only take on the values $\pm 1$, but can vary over the reals.
We choose a linear relation between the two sets of variables
\[
\label{eq:StoRho}
\rho_i = \delta \frac{1+s_i}{2} + b
\]
where $\delta$ and $b$ are constants. From the form of solution in Eq. \ref{eq:Hopfield}
the $s_i$ still must be bounded by $\pm 1$, and therefore $\rho_i$ is bounded to go between $b$ and
$b+\delta$. These bounds are chosen to be biologically sensible, meaning that the unbound
concentrations, $\{\rho_i\}$, need not become arbitrarily small or large for the mechanism proposed here to work.

Corresponding to the learned patterns $t_i^\alpha$ in Eq. \ref{eq:Hebb}, will be the {\em learned unbound
concentrations} defined as
\[
\label{eq:TtoP}
p_i^\alpha = \delta \frac{1+t_i^\alpha}{2} + b
\]
In analogy with learning algorithms, we are storing $M$ patterns of $\rho$, with the $\alpha$th pattern
having unbound concentrations of $\{p_i^\alpha\}_i$.

Similarly, we would like to relate the Boltzmann Machine couplings in Sec. \ref{Sec:RelToBoltzMach}
to the equilibrium constants $K_{ij}$ above, through
\[
\label{eq:JtoK}
K_{ij} = \epsilon \frac{1+J_{ij}}{2} + a
\]
where $a$ and $\epsilon$ are both constants.
If we restrict $|J_{ij}| \le 1$ for all $i$ and $j$, by appropriate normalization of
Eq. \ref{eq:Hebb}, this means that $a < K_{ij} < a+\epsilon$. 
This allows us to choose physically sensible values for the equilibrium constants. 

\subsection{Creation rate}
\label{sec:CreationRate}

To relate the Hopfield model solution Eq. \ref{eq:Hopfield} to the kinetic equation for our
RNA system, we choose the form of $f$ in Eq.\ref{eq:FirstOrderC}. 
% Denoting the ratio $r_i \equiv C_i/\rho_i$, we write
\begin{equation}
\label{eq:FormOfF}
f(C_i,\{\rho_k\}) =   \frac{C_i}{\rho_i} S(\frac{4}{\epsilon}(\frac{C_i}{\rho_i} -1) -2(1+2\frac{a}{\epsilon})\sum_{j=1}^N \rho_j -
 \frac{2(\delta+2b)}{\epsilon}\sum_j K_{ij} + (\frac{2a}{\epsilon}+1)(\delta+2b)N)
\end{equation}
for $i=1,\dots,N$.

The function $S(x)$ is chosen to be
\begin{equation}
\label{eq:Sigmoid}
S(x) = \frac{\delta}{2}[1+\tanh(\beta x/N)] + b
\end{equation}

Eqs. \ref{eq:FirstOrderRho}, \ref{eq:FirstOrderC}, \ref{eq:FormOfF}, and \ref{eq:Sigmoid} along
with arbitrary initial conditions, fully define
the dynamics of the unbound and bound RNA concentrations, given a set of equilibrium constants
$K_{ij}$.

\section{Equivalence of RNA system to machine learning algorithm}
\label{sec:equivRNAtoML}

We are interested in the long time steady state solution, where 
there is no time dependence and therefore all time derivatives are zero.

Eq. \ref{eq:FirstOrderRho} becomes \ref{eq:ConcentrationModification} in this limit, and we can
write
\begin{equation}
\label{eq:KP}
\sum_j K_{ij} \rho_j = \frac{C_i}{\rho_i}-1 . 
\end{equation}

Similarly, Eq. \ref{eq:FirstOrderC} becomes
\begin{equation}
\label{eq:SelfConsistentC}
C_i = f(C_i,\{\rho_k\}) .
\end{equation}

Substituting in Eq. \ref{eq:FormOfF} 
\begin{equation}
\label{eq:FTransformed}
C_i =  \frac{C_i}{\rho_i} S(\frac{4}{\epsilon}(\frac{C_i}{\rho_i} -1) -2(1+2\frac{a}{\epsilon})\sum_{j=1}^N \rho_j -
 \frac{2(\delta+2b)}{\epsilon}\sum_j K_{ij} + (\frac{2a}{\epsilon}+1)(\delta+2b)N)
\end{equation}
Canceling the $C_i$'s 
and using Eqs. \ref{eq:KP}, \ref{eq:StoRho} and \ref{eq:JtoK}, solving for $s_i$, and substituting Eq. \ref{eq:Sigmoid}
finally gives the same form as Eq. \ref{eq:Hopfield}, 
\begin{equation}
\label{eq:HopfieldRNA}
s_i = \tanh(\frac{\beta\delta}{N}\sum_j J_{ij} s_j)
\end{equation}

It is not necessary that a $\tanh$ function be used here. A variety sigmoidal shaped curve should have the correct
properties, with similar efficacy.

The above analysis does not show  that these equation will lead to this steady state solution, 
and indeed, if the time scales are not as described here, it can lead to different steady state
behavior. Next we will explore this problem numerically to find out if the equivalence to
the above solution is viable, and if it does lead to machine learning.

\subsection{Numerical results}
\label{sec:NumResults}

The above model was implemented numerically. We take a system of 
$N=50$ RNA species with unbound concentrations $\{\rho_i\}$ and total concentrations
$\{C_i\}$ as described above, and evolve these over time according to the Eqs.
\ref{eq:FirstOrderRho}, \ref{eq:FirstOrderC}, \ref{eq:FormOfF}, and \ref{eq:Sigmoid}.
The equilibrium constants $K_{ij}$'s were chosen according to Eq. \ref{eq:Hebb} where we
analyzed the retrieval of $M=3$ patterns. The $t_i^\alpha$ were chosen randomly to be $\pm 1$
and these correspond to values of $\rho$ given in Eq. \ref{eq:TtoP}.
After evolution for sufficient time to have converged, we checked to see if the pattern of $\rho$'s found
was one of the three patterns that were encoded in the $K_{ij}$'s.

When the transformation of Eq. \ref{eq:JtoK} was applied, the final $K$'s were scaled so that
their values were between $a$ and $a+\epsilon$. The values of the $\rho$'s were also rescaled
according to Eq. \ref{eq:StoRho}. We found that the values of these rescaling parameters,
$a$, $b$, $\epsilon$ and $\delta$, did not have a strong effect on convergence of the model.

The ratio of the two timescales $\tau_C/\tau_\rho$ needed to be sufficiently
large to obtain consistent convergence over a wide range of initial conditions. 
A ratio of $\tau_C/\tau_\rho = 100$ was found to work in all cases. 
With smaller values, such as  $\tau_C/\tau_\rho = 10$, convergence worked well for
some initial conditions but not for all of them.

When starting with arbitrary initial $\rho_i$'s, 
the corresponding values of $C_i$ were chosen by rearranging Eq. \ref{eq:KP}
\begin{equation}
\label{eq:Ceq=psumK}
C_i = \rho_i (\sum_j K_{ij} \rho_j + 1) 
\end{equation}

The equations were evolved using an explicit embedded Runge-Kutta-Fehlberg 4(5) method, with
a step size of $0.1$.

We investigated several important properties of the system's dynamics, the basin
of attraction starting from $\rho$'s that were different from the initial patterns.
We also investigated how altering the optimal $K_{ij}$'s influenced the final patterns found.
We also investigated how the number of hidden units influence the system's performance.
The following two subsections study sensitivity to deviations in the $\rho$'s and deviations
in the $K$'s. In the third subsection, we study the effects of clamping some of the concentrations
to fixed values, to study how well such systems perform as Boltzmann machines.

\subsubsection{Sensitivity to unbound concentrations}

We tested out the basin of attraction of initial values of the $\rho_i$'s.
Because the $\rho$'s continuously vary in time, to compare the converged
solutions to the binary patterns $t_i^\alpha$, we partitioned the $\rho$'s so that
they corresponded to $-1$ if $\rho_i < b + \delta/2$, at $+1$ otherwise, which
is seen from the mapping between the two systems in Eq. \ref{eq:TtoP}.

An initial pattern was altered from $t_i^\alpha$ so that they differed randomly at $n$ locations.
That is, the Hamming distance was set to $n$. Then the system was evolved from this condition to see
if it would relax back to that same pattern $\{t_i^\alpha\}_i$. 
Note that because of symmetry, both $\{t_i^\alpha\}_i$ and $-\{t_i^\alpha\}_i$ are
possible solutions that should be considered when comparing for convergence. Any deviation from the
trained pattern was considered a mistake.

At every initial Hamming distance $n$, we generated $10$ independent sets of $M$ patterns. 
For each of these sets, we tried starting with $10$ randomly altered patterns that we evolved for each of the $M$ patterns.
Altogether, this represents $300$ separate runs for each Hamming distance studied.

We also tried making two separate kinds of alterations to the $t_i^\alpha$'s, ones that conserve the total
number of $1$'s and $-1$, and ones that allow this number to vary. This becomes an important
distinction later on, when we consider a more universal version of this model.

\begin{figure}[htb]
\begin{center}
(a) \includegraphics[width=0.45\hsize]{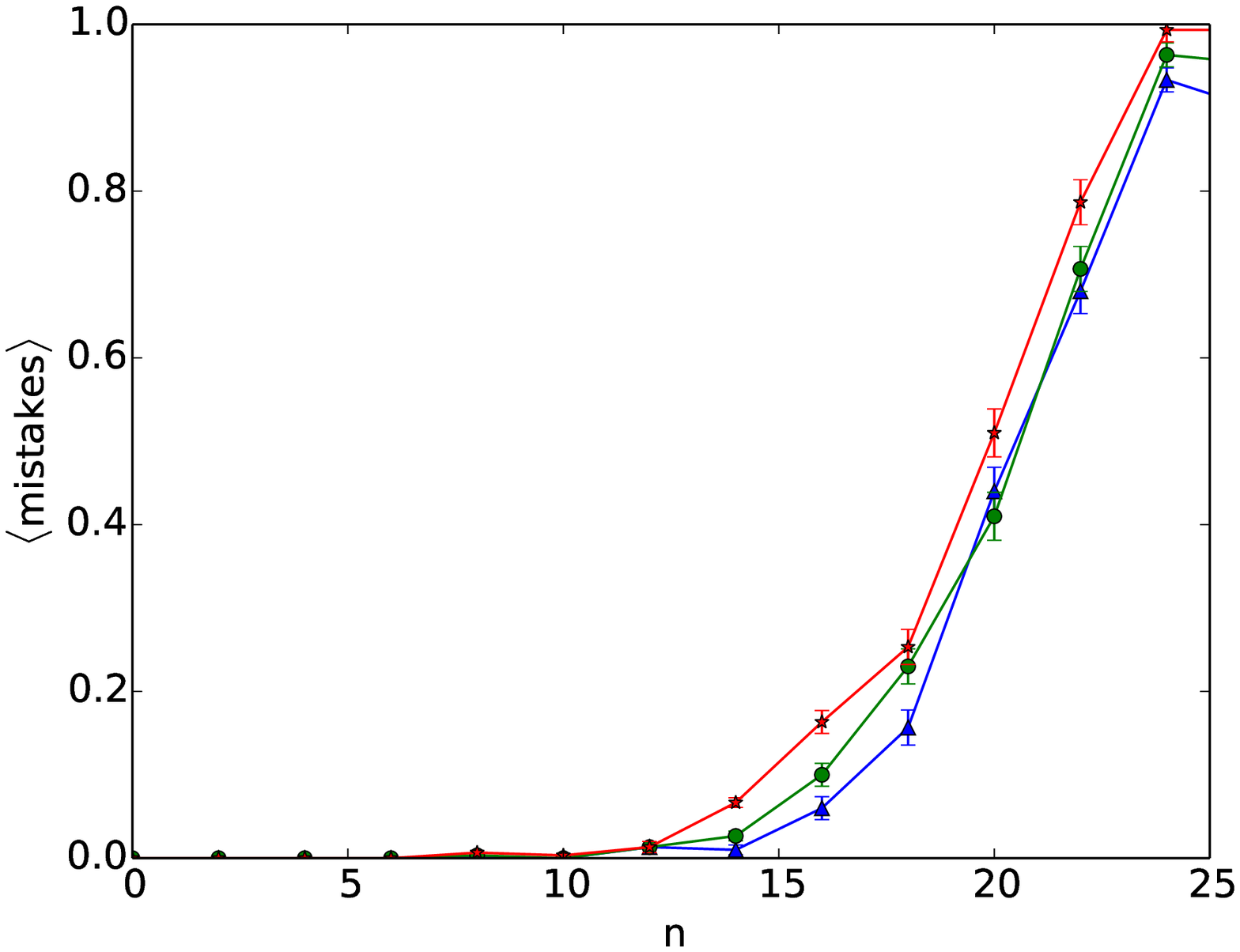} 
(b) \includegraphics[width=0.45\hsize]{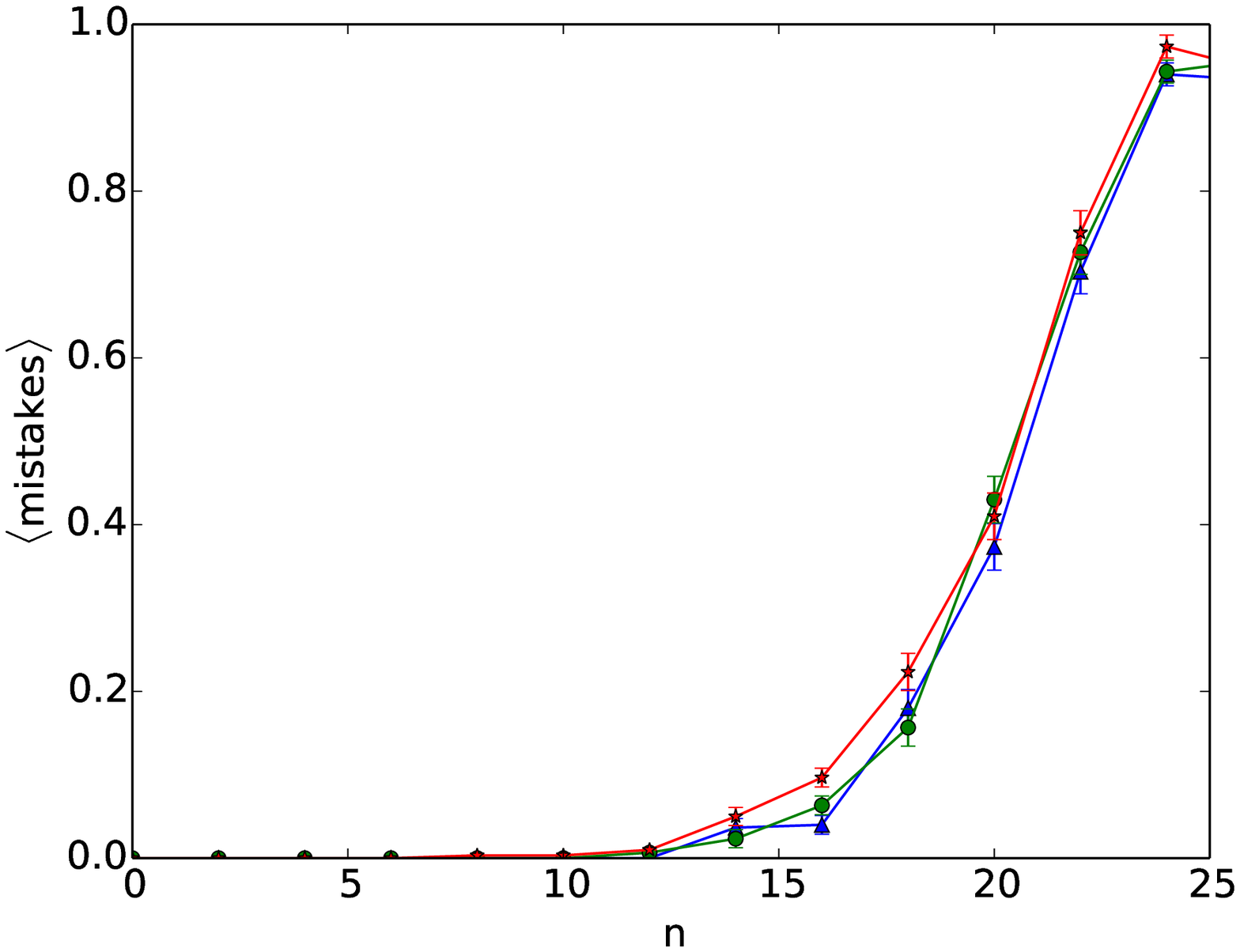} 
\caption
{ 
(a)
The average fractional number of mistakes made as a function of the Hamming distance, $n$, between
the initial state and the pattern. Here the total number of RNA species is $N=50$ and 
the total number of patterns to be recalled is $M=3$.
The triangles have $\beta\delta/N = 0.5$ the circles, $\beta\delta/N = 2$, and
the stars show $\beta\delta/N = 8$. The lines are simply a guide for the eye.
(b) The same case as in (a) except that the total unbound concentration of the initial
state is unchanged.
}
\label{fig:mistakes_vs_hamming}
\end{center}
\end{figure}

We plot the fraction of mistakes as a function of the Hamming distance
cutoff in Fig. \ref{fig:mistakes_vs_hamming}(a). The three graphs show the results for
different values of $\beta$, $\beta\delta/N = 0.5$, $2$, and $8$. With less accurate iteration
methods, it was found that $\beta\delta/N < 8$ was not stable. However with this Runge Kutta
method, the differences between the results are fairly minor. The precise sigmoidal
shape $S(x)$ in Eq. \ref{eq:Sigmoid} is clearly not important.

In these plots, $a=b=0.4$, and $\epsilon=\delta=0.6$ (see Eqs. \ref{eq:StoRho} and \ref{eq:JtoK}).
The results are quite insensitive to these, for example 
$a=b=0.1$, and $\epsilon=\delta=0.8$ yields similar results.

We now consider mutations that do not change the total unbound concentration. Changes to the
patterns are made in random pairs $i$, $j$, so that the sum of the $t_i^\alpha + t_j^\alpha$
stays constant. The results are shown in Fig. \ref{fig:mistakes_vs_hamming}(b). 

\subsubsection{Sensitivity to equilibrium constants}

We now consider the effect that changing the $K_{ij}$'s has on the patterns that are
retrieved. With the usual Hopfield model, it is known to be quite robust
to changes of the connectivity strength, which greatly contrasts with usual digital architecture.
We investigate to what extent this still carries over here.

\begin{figure}[htb]
\begin{center}
\includegraphics[width=0.5\hsize]{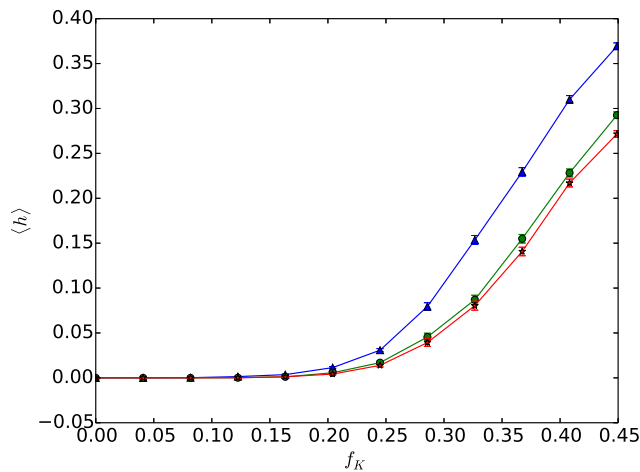}
\caption
{ 
The average normalized Hamming distance is plotted versus fractional mutation frequency of equilibrium constants $f_k$. 
Triangles correspond to $\beta\delta/N = 0.5$, circles $\beta\delta/N = 2.0$, and stars $\beta\delta/N = 4.0$. 
}
\label{fig:mistakes_vs_K}
\end{center}
\end{figure}

We chose
random $i$'s and $j's$ and mutated them by taking $K_{ij} \rightarrow 1 - K_{ij}$,
maintaining symmetry of the matrix. (Recall that
$0 \le K_{ij}\le 1$). We measured the average Hamming distance $H(\{t_i^\alpha\},\{\rho_i\}_i)$,
by translating the $\rho$'s into corresponding discrete spin variables. This way, we are
counting the number of differences between the $t_i^\alpha$'s and the final
pattern, $s_{final,i}$ that emerged. We normalize the Hamming distance by dividing by $N$, 
that is $h = H(\{t_i^\alpha\},\{s_{final,i}\})/N$.
This is computed as a function of the number of
mutations $n_K$ made to the $K_{ij}$'s. To normalize this, we define 
\[
f_K \equiv \frac{2 n_K}{N(N-1)}
\]
We performed these mutations with all other parameters identical to the ones used above, and the results
are shown in Fig. \ref{fig:mistakes_vs_K}. 

\subsubsection{Clamping input concentrations}

Now we consider clamping,  $N_f$ of the $C_i$'s so that they are fixed to predetermined values as the
biochemical network evolves in time, to see if it can correctly associate those clamped inputs to
outputs. This the kind of task that is performed by a
Boltzmann machine. Fig. \ref{fig:BM} illustrates this process. The filled black circles represent the
clamped inputs, and are coupled through the $K_{ij}$'s, represented by lines, to all other units. A single unit, $i$, represents
the concentrations $\rho_i$ and $C_i$. The unfilled circles have $\rho_i$'s and $C_i$'s that vary, and two of these
circles represent outputs. To test out this capability, we start off as before, with $N=50$ and choosing
$M=3$ separate random patterns $S_i = t_i^\alpha$ which are then translated into $\rho_i$'s, again using
Eq. \ref{eq:TtoP}. The couplings $K_{ij}$ are chosen as before as well. 

We have chosen $N_f$ of the $C_i$'s, for $i=N_v+1,\dots,N$, to fixed values, and then let
the other $N_v$ units vary as before, as described by Eq. \ref{eq:FirstOrderC}. We also alter the
initial conditions, by randomly scrambling the remaining values of $\rho_i$, $i=1,\dots,N_v$. 
As usual, the corresponding initial $C_i$'s were chosen through Eq. \ref{eq:Ceq=psumK}.

Out of the $N_v = N-N_f$ $\rho_i$'s that vary, we will regard two of these as output units and
the rest as hidden. We would like to know how well, given the fixed inputs, the system evolves
to finally recall these two output units.

\begin{figure}[htb]
\begin{center}
\includegraphics[width=0.5\hsize]{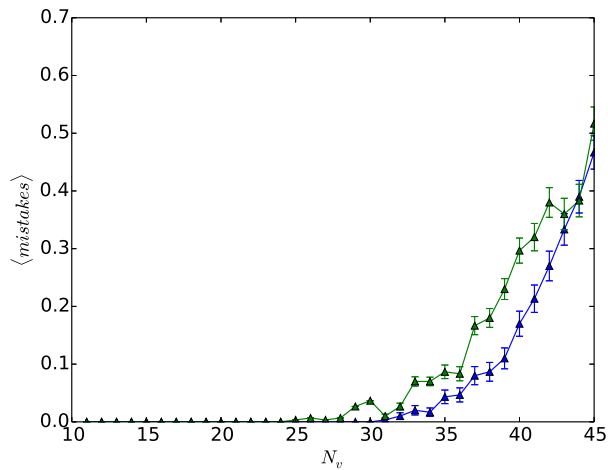}
\caption
{ 
The average fractional number of mistakes made as a function of the number of variable units, $N_v$,
in the Boltzmann machine analog illustrated in Fig. \ref{fig:BM}. The triangles are for the case
$\beta\delta/N = 0.5$, and the stars are for $\beta\delta/N = 4.0$.
}
\label{fig:mistakes_vs_vary}
\end{center}
\end{figure}

Fig. \ref{fig:mistakes_vs_vary} shows the fractional number of mistakes plotted versus the number of
variable units, $N_v$, for two different temperatures, $\beta\delta/N = 0.5$ and $4.0$.

\section{More universal regulation}
\label{sec:MoreUnivReg}

The numerical results of the previous section illustrate that for a wide range of parameters, this genetic biochemical network
has capabilities quite similar to those of power machine learning algorithms. The main criticism of
this system is the rather contrived nature of the function $f$ in Eq. \ref{eq:FormOfF}. This
complicated form was designed to give the same steady state solutions as the analogous machine
learning system. But it is not clear how this could be implemented biologically. We now show how
this mechanism can be greatly simplified, leading to a much stronger case for biological 
relevance.

We start by writing Eq. \ref{eq:FormOfF} as 
\begin{equation}
\label{eq:rewriteFormOfFAcal}
f(C_i,\{\rho_k\}) =   \frac{C_i}{\rho_i} S(\frac{4}{\epsilon}(\frac{C_i}{\rho_i} -1) - {\cal A})
\end{equation}
with
\begin{equation}
\label{eq:FormOfcalA}
{\cal A} = 2(1+2\frac{a}{\epsilon})\sum_{j=1}^N \rho_j +
 \frac{2(\delta+2b)}{\epsilon}\sum_j K_{ij} + (\frac{2a}{\epsilon}+1)(\delta+2b)N)
\end{equation}

The complication here is that $\cal A$ is not a constant but depends on unbound densities $\rho_i$.
However, it 
only depends on the sum of all of these. We start by considering initial conditions that
still differ from the patterns $p_i^\alpha$ but have the same total sum. We ask if replacing
$\sum \rho_i$ by a constant value will influence the steady state, that is, Eq. \ref{eq:HopfieldRNA}.
Therefore we will define a more ``universal" creation function as follows
\begin{equation}
\label{eq:FormOfFU}
f(C_i,\{\rho_k\}) =   \frac{C_i}{\rho_i} S(\frac{4}{\epsilon}(\frac{C_i}{\rho_i} -1) 
-2(1+2\frac{a}{\epsilon})\sum_{j=1}^N p_j^\alpha -
\frac{2(\delta+2b)}{\epsilon}\sum_j K_{ij} + (\frac{2a}{\epsilon}+1)(\delta+2b)N)
\end{equation}
where we have replaced the sum of the $\rho_i$'s by a sum over pattern $\alpha$, $p_i^\alpha$.
We can now follow the same steps as we employed in Sec. \ref{sec:equivRNAtoML}
to relate this biochemical system to the machine learning spin system.
In this case, Eq. \ref{eq:FTransformed} now has the term in the argument of the function $S$,
$2(1+2\frac{a}{\epsilon})\sum_{j=1}^N \rho_j$, replaced by
$2(1+2\frac{a}{\epsilon})\sum_{j=1}^N p_j^\alpha$. The argument of $S$ now
differs from its previous value by 
\[
\label{eq:Deltah}
\Delta_h \equiv
2(1+2\frac{a}{\epsilon})\sum_{j=1}^N (p_j^\alpha-\rho_j)
= 
(1+2\frac{a}{\epsilon})\delta \sum_{j=1}^N (t_j^\alpha-s_j)
\]
where in the last equality, Eqs. \ref{eq:TtoP} and \ref{eq:StoRho} have allowed us 
to translate this difference into spin variables.
To determine the effect of this term, for simplicity, we will consider the case where $\sum
t_i^\alpha = 0$. Because these patterns are chosen at random, for large $N$, this is a reasonable
assumption. The simplest way that the mean field solutions in Eq. \ref{eq:Hopfield} or equivalently
Eq. \ref{eq:HopfieldRNA} can be understood\cite{HertzKroghPalmer} is by utilizing the random nature of the $t_i^\alpha$'s. 
Examining the argument on the right hand side of this equation, and using Eq. \ref{eq:Hebb}
\[
\sum_{i=1}^N J_{ij} s_i = {\cal N} \sum_{i=1}^N \sum_{\beta=1}^M t_i^\beta t_j^\beta s_i.
\]
Previously we chose $|J_{ij}| \le 1$. 
Because the patterns are independent, to achieve this, the normalization factor $\cal N$, will be of order $1/sqrt{M}$ (with
additional logarithmic corrections that are not important to our conclusions).
If we change ``gauge", writing $\sigma_i \equiv s_i t_i^\alpha$, 
\[
\label{eq:CmpreHopfieldTerms}
\sum_{i=1}^N J_{ij} s_i 
= {\cal N} [\sum_{i=1}^N   t_j^\alpha \sigma_i + 
\sum_{\beta\ne \alpha}^M t_j^\beta \sum_{i=1}^N  t_i^\beta t_i^\alpha  \sigma_i]
= {\cal N} [N  t_j^\alpha \sigma_i + 
\sum_{\beta\ne \alpha}^M t_j^\beta \sum_{i=1}^N  t_i^\beta t_i^\alpha  \sigma_i]
\]
If we choose the pattern $s_i = t_i^\alpha$, then $\sigma_i = 1$ for all $i$. The
first term in the last equality is $({\cal N} N) t_j^\alpha$. Because the patterns
are random, the second term has a term of order $\pm {\cal N} \sqrt{N M}$, which for $N \gg M$, is
negligible compared to the first term. In this limit, all of the other patterns $\beta \ne \alpha$
can be ignored, and we have the mean field equation for a single pattern, the so-called ``Mattis
model"~\cite{amit1985spin}. As is evident from Eq. \ref{eq:HopfieldRNA}, this is satisfied, and
using the normalization factor ${\cal N} \propto 1/\sqrt{M}$, 
the size of the dominant term is of order $N/\sqrt{M}$. 

Now we return to the corrections to  Eq. \ref{eq:HopfieldRNA} given by Eq. \ref{eq:Deltah}.
Doing the same kind of estimation, $\Delta_h$ has magnitude $\pm \sqrt{N}$. Therefore
comparing the factors of $N$ and $M$ with Eq. \ref{eq:CmpreHopfieldTerms}, which is of
order $N/\sqrt{M}$, $\Delta_h$ is  negligible for $N \gg M$, which is the case that 
we are already considering.

Therefore for the random patterns (typically used in machine learning
problems such as the Hopfield model) with a constant sum, and for large $N$, replacing the summation of the $\rho$'s by
Eq. \ref{eq:FormOfFU} is not expected to alter the steady state solutions of Eq.
\ref{eq:HopfieldRNA}.

We will therefore only consider learnt patterns with the property that $\sum_i p_i^\alpha$ is
a constant and does not depend on $\alpha$. In the most important 
case\footnote{Analysis of the Hopfield model for nonzero $\sum t_i^\alpha$ can also be performed~\cite{amit1987information}.}
considered above, 
this corresponds to a machine learning
problem with $\sum_i t_i^\alpha = 0$ for all $\alpha = 1,\dots,M$. 
We therefore can write
\[
\label{eq:Ptot}
P_{tot} \equiv \sum_i p_i^\alpha = N (\frac{\delta}{2}+b)
\]
where the last equality used Eq. \ref{eq:StoRho}.

This shows that we can modify our  creation mechanism so that it only depends on the ratio $\frac{C_i}{\rho_i}$
\begin{equation}
\label{eq:rewriteFormOfFwithA}
f(C_i,\{\rho_k\}) =   \frac{C_i}{\rho_i} S(\frac{4}{\epsilon}(\frac{C_i}{\rho_i} -1) - A)
\end{equation}
where $A$ maintains a constant value in time and only depends on the fixed parameters, such as the
equilibrium constants $K_{ij}$ and the total sum of learned pattern concentration $P_{tot}$,
\begin{equation}
\label{eq:FormOfA}
 A = 2(1+2\frac{a}{\epsilon})P_{tot} +
 \frac{2(\delta+2b)}{\epsilon}\sum_j K_{ij} + (\frac{2a}{\epsilon}+1)(\delta+2b)N)
\end{equation}

This approach assumes that $\sum_i \rho_i$ starts close to $P_{tot}$. If it does not, then
an additional regulatory mechanism is needed to drive this sum towards $P_{tot}$. This would operate
in a similar way to other homeostatic mechanisms. If $P_{tot}$ deviates, the
total RNA concentration should vary as well. Mechanisms would need need to
ensure that this stays at a well defined value. But even if we completely ignore
such a general mechanism, we can study numerically compared to our previous
results. We will see that it still works surprisingly well.

Eqs. \ref{eq:FirstOrderC}, \ref{eq:rewriteFormOfFwithA} and \ref{eq:FirstOrderRho} define
the system of equations to be evolved in time. The creation of RNA is now much simpler to
describe. It depends on the ratio of total concentration to unbound concentration.

This model will now be investigated numerically.

\subsection{Numerical results}
\label{sec:UnivNumResults}

The above model with this much simpler creation function, was studied using the same parameters
as in Sec. \ref{sec:NumResults}, e.g. $M=3$, and $N=50$. 
As mentioned above, we are assuming a general homeostatic mechanism,
as considered earlier in Fig. \ref{fig:mistakes_vs_hamming}(b) where
the initial unbound concentrations preserve their total value, $\sum_i \rho_i$.

\begin{figure}[htb]
\begin{center}
\includegraphics[width=0.5\hsize]{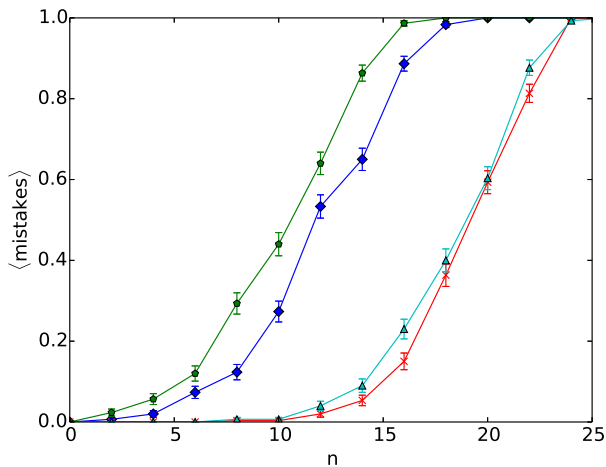}
\caption
{ 
The average fractional number of mistakes made as a function of the Hamming distance between
the initial state and the pattern, $n$. Here the total number of RNA species is $N=50$ and 
the total number of patterns to be recalled is $M=3$.
The triangles are the results for $\beta\delta/N = 4$ and the initial total unbound concentration
equal to that of the patterns to be recalled. The pentagons are for the same parameters but
the initial unbound concentration has one more up ``spin".  The crosses 
are with those same initial conditions but with $\beta\delta/N = 8$, and the diamonds are
with one more up ``spin". The lines are simply a guide for the eye.
}
\label{fig:mistakes_vs_hamming_fixed}
\end{center}
\end{figure}

To investigate how well this system works in more detail, we consider systems that are regulated
so that the total concentration of unbound RNA starts off close to $P_{tot}$, but is otherwise
scrambled. This was done
with the same procedure as in Fig. \ref{fig:mistakes_vs_hamming}(b). The four graphs in Fig. \ref{fig:mistakes_vs_hamming_fixed} show
the fractional number of mistakes as a function of the Hamming distance, $n$, between the initial
state and the pattern to be recalled. The recall works best when the total unbound concentration is
maintained at the correct final amount, and is less good when it deviates from that. This shows the
necessity for carefully regulating the total RNA concentrations.

\begin{figure}[htb]
\begin{center}
\includegraphics[width=0.5\hsize]{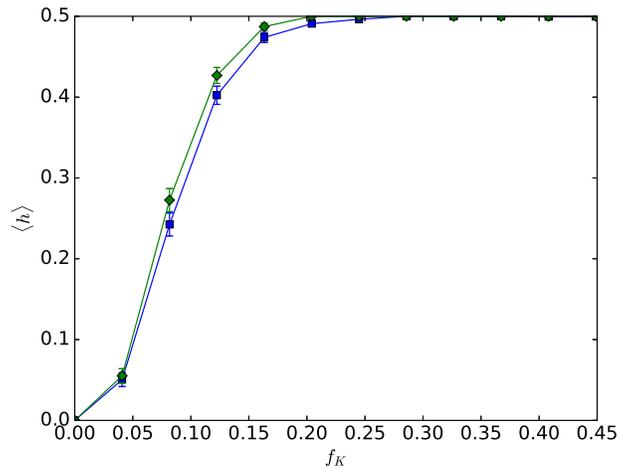}
\caption
{ 
The average normalized Hamming distance is plotted versus fractional mutation frequency, $f_K$, of equilibrium constants $K_{ij}$. 
Diamonds correspond to $\beta\delta/N = 4.0$, and squares $\beta\delta/N = 8.0$. 
}
\label{fig:mistakes_vs_K_fixed}
\end{center}
\end{figure}

Similarly, the variation of the normalized Hamming distance as a function of the mutation frequency $K_{ij}$'s is
shown in Fig. \ref{fig:mistakes_vs_K_fixed}. In comparison with Fig. \ref{fig:mistakes_vs_K} it shows
more sensitivity. This is not surprising, because the sum over the $K_{ij}$'s in Eq. \ref{eq:FormOfFU}
will no longer be the same, and this kind of variation was not taken into account in the analysis of
the last section, only changes in the $\rho_i$'s. Biochemical circuitry could be posited to
further adjust $A$, but because mutations in the $K_{ij}$'s happen in the process of
evolution, additional biochemical circuitry is not necessary
if we allow changes in the value of $A$ to occur during evolution. Any detailed
discussion on this topic becomes far too speculative to warrant serious consideration
at this stage.

\begin{figure}[htb]
\begin{center}
\includegraphics[width=0.5\hsize]{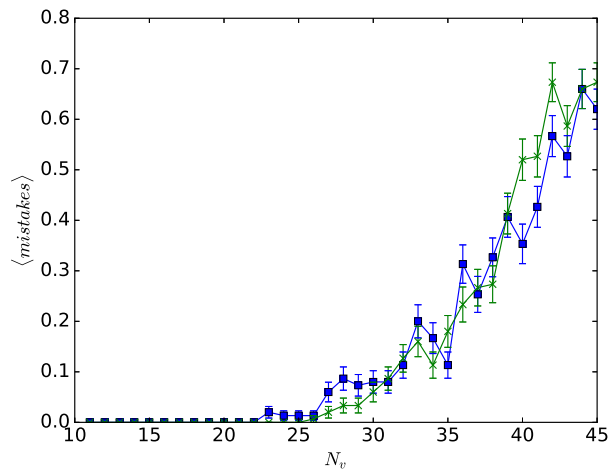}
\caption
{ 
The average fractional number of mistakes made as a function of the number of variable units, $N_v$,
in the Boltzmann machine analog illustrated in Fig. \ref{fig:BM}. The crosses are for the case
$\beta\delta/N = 0.5$, and the squares are for $\beta\delta/N = 4.0$.
}
\label{fig:mistakes_vs_vary_fixed}
\end{center}
\end{figure}

In reference to Boltzmann machines, Fig. \ref{fig:mistakes_vs_vary_fixed} shows the fractional number of mistakes plotted versus the number of
variable units, $N_v$, for two different temperatures, $\beta\delta/N = 0.5$ and $4.0$. It is quite
similar to Fig. \ref{fig:mistakes_vs_vary}.

\section{Discussion}
\label{sec:discussion}

\subsection{Magnitude of RNA interactions}
\label{sec:MagRNAint}

We can estimate the density and timescales for non coding RNA from the available data.

We expect that little or none of the long ncRNA in the cytoplasm will import back into the
nucleus and therefore we will confine our attention to RNA that is preferentially localized
to the nucleus. Given its regulatory role, it is not surprising
that ncRNA is, on average, preferentially enriched in the nucleus, in contrast with mRNA that is exported into the
cytoplasm. The ratio of nuclear to cytoplasmic ncRNA 
is approximately 1~\cite{derrien2012gencode}. In other words, that approximately half of it is localized to the
nucleus. 

Given its predominantly regulatory function, it is not surprising that the total
amount of ncRNA present in a cell is estimated to be lower than the
total amount of mRNA. It is appears, on average that long ncRNA has
approximately a tenth of the abundance of mRNA, although this number
fluctuates substantially depending cell type, much more than for
mRNA~\cite{cabili2011integrative}. The total number of mRNA in a mammalian
cell is approximately~\cite{milo2015cell} $5\times 10^5$. This implies that the total amount
of long ncRNA is approximately $N_L = 2\times 10^4$ per cell nucleus.

A mammalian cell nucleus has a radius of approximately $r_n = 3 \mu m$. This gives a 
long ncRNA density of $\rho_L = 3 N_L /(4\pi r_n^3)$, or an average separation of
$r_L = \rho_L^{-1/3}$, which is approximately $0.18 \mu m$. Because of the heterogeneous nature of the
nuclear environment, it is not easy to get a precise estimate for diffusion coefficients,
but mRNA in the nucleus appears to have a diffusion coefficient $D \approx 0.1 \mu m^2/s$~\cite{politz2000movement}
which should be similar to that of ncRNA, although there will be a large range depending
on the species.  Therefore the time for an ncRNA molecule to move a distance $r_L$ is on average $r_L^2/6D$, which is
approximately, $.05 s$. In that time, it will not necessarily encounter another ncRNA molecule, and this will
increase the time scale by a factor of $r_L/d$, where $d$ is a measure of the size of the molecule. 
For a $1000$ base pair RNA, with a persistence of length of approximately $40$ {{\AA}}, this gives $d \approx 40 nm$,
and therefore $r_L/d \approx 5$. This means that the collision time is approximately $0.25 s$. 

The half-life for ncRNA in the nucleus is of order 30 minutes to an hour~\cite{lee2012epigenetic}. Therefore
we expect that  there is a
large difference between the time scale for equilibration and for degradation of these ncRNA molecules,
as assumed by the model here.

\subsection{Mechanisms for creation}
\label{sec:MechForCreation}

Here we described a biophysical mechanism capable of performing sophisticated computation, using
RNA produced by the genome. The ability to make high-level decisions based on its inputs has obvious
advantages to a biological organism. Taking advantage of the large amounts of non-coding RNA
produced by a cell, should increase the computational capabilities roughly according to the number
of mutual interactions between the RNA. For example, if a mechanism similar to this was to be utilized, it
would allow an organism to learn from its previous evolutionary history by encoding past
environments in the values of the equilibrium constants $\{K_{ij}\}$. However, it is far
from clear that a mechanism similar to what has been described here, does in fact exists.

In this section we describe some potential ways that 
Eq. \ref{eq:rewriteFormOfFwithA}, which gives the rate of creation of a RNA species $i$, 
could be realized in practice. The crucial quantity that the system must measure is the
fraction of unbound RNA. Fig. \ref{fig:creation_function} plots the general form of this
function. The specific parameters used here are $\delta = 0.4$ and $b = 0.6$. The curve
starts at $C/\rho = 1$ and increases from there. This creation rate deviates subtly
from linearity, and a large number of functional forms with this general shape would
be suitable. For example we have already seen that $\beta$ can be considerably altered
with only a modest effect on performance. The choice of the $\tanh$ function was also
not necessary and a wide variety of sigmoidal shaped functions are expected also work
as has been investigated in neural network models~\cite{HertzKroghPalmer}.
We will now discuss what kinds of models could be expected to give this general
shape for the creation rate.

{\begin{figure}[htb]

\begin{center}
\includegraphics[width=0.5\hsize]{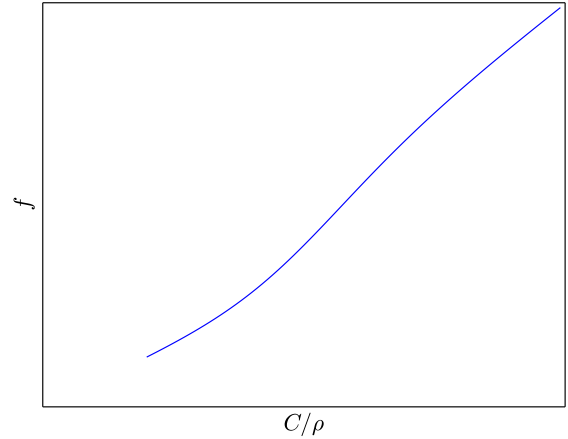}
\caption
{ 
The creation rate of RNA given by Eq. \ref{eq:rewriteFormOfFwithA} as a function of the total to unbound RNA
$C/\rho$, for a given RNA species.
}
\label{fig:creation_function}
\end{center}
\end{figure}

We first discuss a non-genomic mechanism, which is the most direct, but least likely to exist biologically.
There is evidence that that there is a biochemical pathway recreating a similar function
as  RNA-dependent RNA polymerase (RdRP)~\cite{spiegelman1965synthesis}, in humans~\cite{kapranov2010new}.
RNA molecules could in principle
be copied without reference to DNA. But in this case, the rate of transcription should depend on the 
ratio of $C/\rho$, that is, the total to unbound RNA of a single species. One mechanism to do this would be
to have a double stranded RNA sensor. Toll-like~\cite{akira2001toll} double stranded RNA sensors do exist,
such as TLR3~\cite{alexopoulou2001recognition} but these are membrane spanning however.
This possibility is shown pictorial in Fig. \ref{fig:RdRPcreation}(a). A less fanciful mechanism is illustrated in 
Fig. \ref{fig:RdRPcreation}(b). Here the putative RdRP is regulated by a site on it, shown
as a circle. If RNA binds to this site, it will inhibit transcription. The most likely RNA to be bound
is the same species that is being copied due to its close proximity to the RdRP. This potential
binding process is shown by the arrow going from the end of the transcribed RNA, and pointing to the
binding site. If a third RNA molecule, shown in light grey, associates with the copied RNA, it will
inhibit binding to this site. This will give enhanced RNA creation as the ratio of total to unbound
RNA increases.

\begin{figure}[htb]
\begin{center}
(a)
\includegraphics[width=0.45\hsize]{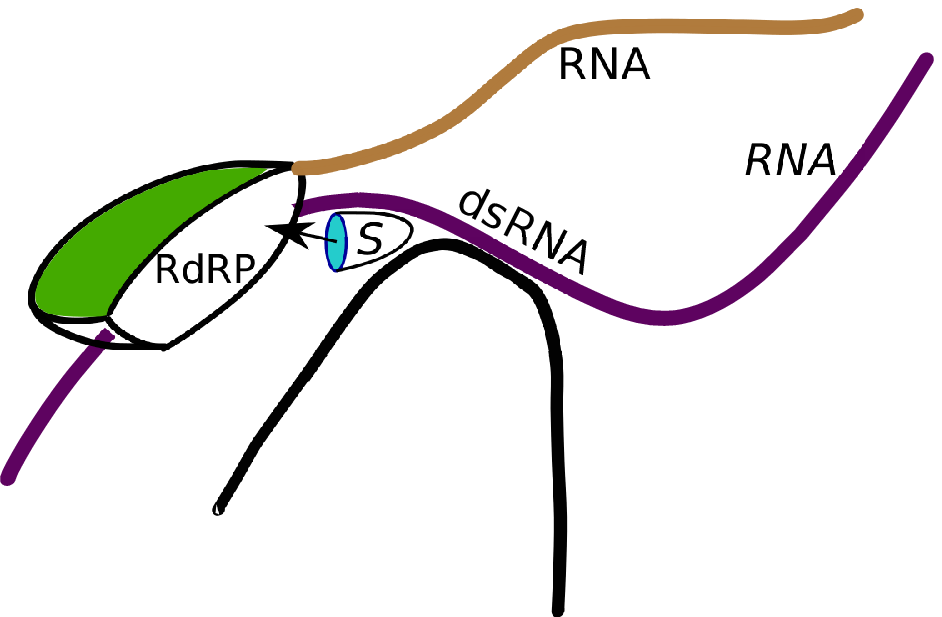}
(b)
\includegraphics[width=0.45\hsize]{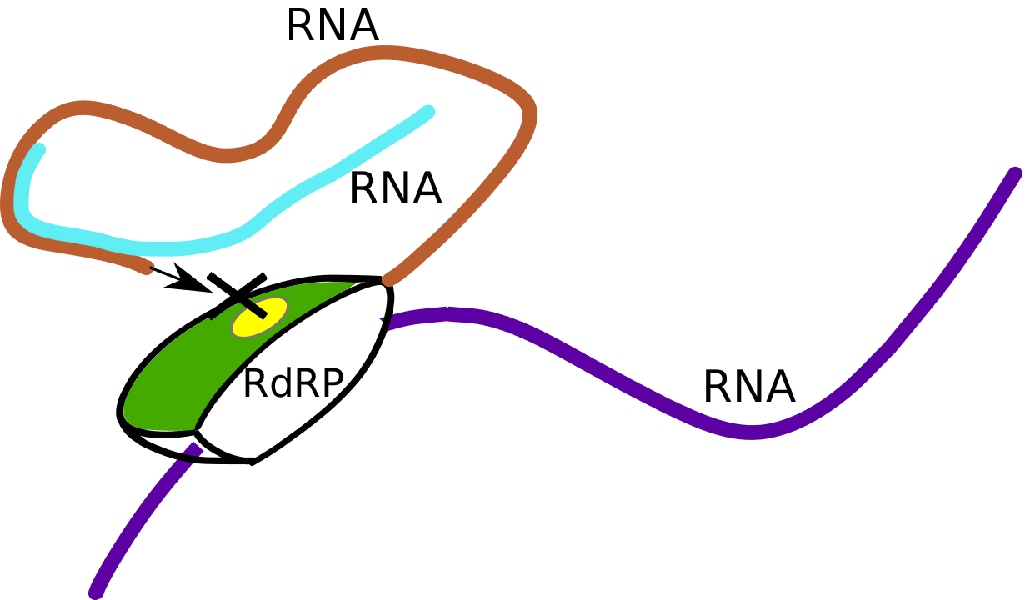}
\caption
{ 
Two possible mechanisms for enhanced RNA transcription when additional RNA has been bound.
(a) A RNA-dependent RNA polymerase (RdRP) copies an RNA molecule that is bound to another one forming a
region of double stranded RNA (dsRNA). An RNA sensor, S, detects the dsRNA which then regulates the
rate of transcription of the RdRP. (b) A site on RdRP, shown by the circle, represses transcription
when RNA is bound to it. This is indicated by the arrow pointing from the RNA end to the binding site. 
The transcribed RNA is inhibited from associating with this regulatory site, by binding to another
RNA molecule, shown in light grey.
}
\label{fig:RdRPcreation}
\end{center}
\end{figure}

We now consider more conventional and promising genomic mechanisms that give creation
rates similar to Fig. \ref{fig:creation_function}.
There are a number of theoretical possibilities for how the creation of RNA can depend on the
ratio of total to unbound RNA.
The unbound RNA can interfere~\cite{agrawal2003rna}, with the translation of an activator protein
specific to the RNA species being transcribed from the DNA. The larger the amount of free RNA, the lower the
rate activator production, and hence the lower the rate of RNA production. 

A more concrete possibility is a similar mechanism that is known to operate in 
some situations~\cite{sigova2015transcription,takemata2016local,takemata2017role}.
Many ncRNAs are transcribed around DNA regulatory elements, such as enhancers.
These ncRNA appear to increase the binding of transcription factors, which increases
transcription. For example, in fission yeast {\em Schizosaccharomyces pombe}, transcription
of ncRNA by RNA polymerase II (RNAP II), from the promoter region of $fbp1^+$ 
(fructose-1,6-bisphosphatase 1), has been shown to depend on the amount of that ncRNA
that is present. The reason for this is due to the ability of this ncRNA to facilitate
the binding of a transcription factor Atf1 on the $fbp1^+$ promotor~\cite{takemata2016local}.
The mechanism for this has been hypothesized to be due to its ability of the ncRNA to down-regulate~\cite{takemata2016local}
corepressor functions of Tup proteins~\cite{mukai1999conservation,asada2015antagonistic}.

Another example of the above enhancement is work in embryonic stem cells of the transcription factor 
Ying Yang 1 (YY1)~\cite{sigova2015transcription}. A number of pieces of evidence pointed to similar
enhanced transcription in the presence of ncRNA. For example, artificially tethering RNA near YY1 binding
sites, increased YY1 occupancy. These results suggest that ncRNA that is transcribed in proximity to
YY1 acts to enhance further ncRNA transcription in this region. 

Various models have been proposed~\cite{takemata2017role} on how this enhancement could take place,
including the trapping of transcription factors by the ncRNA, the recruitment of proteins that increase
transcription factor binding, and the inhibition of proteins the repress transcription factor binding.
These mechanisms are quite general, they imply that an increase in ncRNA concentration around some particular
regulatory elements, should enhance further ncRNA transcription. This increased activation by ncRNA is
a known general function of it~\cite{lee2012epigenetic}.

In the case studied here, this mechanism can potentially lead to the desired behavior shown in
Fig. \ref{fig:creation_function}. Binding of additional ncRNA will increase the local ncRNA concentration
which, as argued above, will lead to an increased rate of ncRNA transcription. This is illustrated in Fig.
\ref{fig:RNAPcreation}. RNA polymerase (labeled RNAP), transcribes ncRNA from DNA. This ncRNA enhances the binding of
a transcription factor (TF). Binding of more ncRNA will further increase transcription due to the increased
concentration of ncRNA near TF. Note that this mechanism is measuring bound ncRNA, rather
than measuring unbound ncRNA. But it is measuring the probability of binding for an individual ncRNA
molecule that is being transcribed. This gives a measure of precisely what we want, the ratio of
bound ncRNA to its total concentration for species $i$, which equals $1-\rho_i/C_i$. The fact that
it correctly divides $\rho_i$ by the total concentration, and is similar to known enhancement
mechanisms, makes it a promising direction to consider.

{\begin{figure}[htb]
\begin{center}
\includegraphics[width=0.5\hsize]{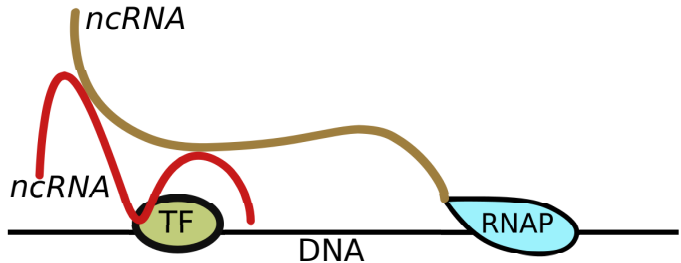}
\caption
{ 
Possible mechanism for enhanced transcription due to bound RNA. RNA polymerase II (RNAP) transcribes
DNA producing ncRNA. The presence of RNA enhances the binding of a transcription factor
(TF), further enhancing the transcription of ncRNA.
}
\label{fig:RNAPcreation}
\end{center}
\end{figure}

\section{Conclusions}

Genetic networks are extremely complex and individual pathways have taken years of
study to elucidate. It is quite apparent by now that ncRNA plays an important role
and is not just ``junk" as had been previously hypothesized~\cite{MercerDingerMattick,ENCODE,djebali2012landscape}.
The purpose of this work is not hypothesize yet another theoretical model that can be added to the 
list of potential mechanisms that biology may be using. Instead, it is to take a step back and
look for new paradigms that can be used to understand genetic regulation. 

Instead of thinking of interactions between individual elements as having a substantial
and measurable effect on each other's behavior, the view taken here, is that there is a class of
interactions that are negligible, but collectively they are able to control behavior, perhaps even
more effectively, than the sparse network models currently employed. Of course there
are many examples where a single interaction has a large effect on gene expression, however
here we examined if collective regulation could also play a substantial role.  

Collective regulation is certainly
possible mathematically, and is the same architecture that is now used in machine learning systems~\cite{HertzKroghPalmer}.
What we have argued here is that this is also biologically and physically plausible.
It is certainly the case that there are strong specific interactions that are
involved in many regulatory pathways. However there is also a large amount of ncRNA that is highly associative,
and is not very specific. The approach taken here is to accept the existence of
thousands (or millions) of potential interactions between different RNA species and understand how these
could have evolved from junk inserted by retroviruses to useful additions to the cell's genome. 

The interactions between the RNA molecules in equilibrium yield a chemical equilibration
between bound and unbound states. In reality there will be many higher order interactions
and different internal states of molecules. These will surely affect the computation
of these systems, but would not necessarily diminish their computational capabilities. Similarly,
models of neurons that only consider two-body interactions, such as Boltzmann machines,
leave out a lot of higher body effects that are present with real neurons.

The chemical equilibration formula, Eq. \ref{eq:ConcentrationModification}, contains the seed of how this system 
is related to artificial neural network models, by making an analogy with 
Eq. \ref{eq:SpinH}. The equilibrium constants $K_{ij}$
are analogous to the interactions $J_{ij}$ between different ``spins" in neural networks. It is the
sum of all unbound RNA, weighted with equilibrium constants, that self consistently must
give back the correct concentration of unbound RNA.

The constant degradation of RNA can be taken into account with a first order reaction rate
equation, Eq. \ref{eq:FirstOrderRho}. But there also needs to be a mechanism for
the replenishment of RNA.
First a homeostatic mechanism needs to be included to regulate the total concentration of all RNA.
But the most difficult part of our analysis, that we investigated analytically and numerically, 
is that the creation rate of the {\em ith} species only be a function of the ratio of  
that RNA's total concentration to the amount that is unbound, $C_i/\rho_i$. Such a function should
look similar to what is shown in Fig. \ref{fig:creation_function}}. We were able to show
that this function
can be universal, in that it only depends $C_i/\rho_i$, and with no dependence on the particular
species of RNA being created. 

The physical binding and unbinding of ncRNA should happen according to estimates using
empirical data discussed in Sec. \ref{sec:MagRNAint}. RNA species are also created through a variety of regulatory mechanisms. 
The validity of the proposal outlined here then
boils down to whether there exists RNA creation rates in the nucleus that depend on $C_i/\rho_i$
according to Fig. \ref{fig:creation_function}. We argued in Sec. \ref{sec:MechForCreation}
that in fact, there is evidence for similar mechanisms already. This does not prove
the existence of the kind of computational scheme given here, but argues that it is at least plausible.

These kinds of computational paradigms have several advantages, one being that they
are far more robust than
circuitry with few connections~\cite{HertzKroghPalmer}, and this would mean that
we would expect ncRNA would have a much higher mutation rate than mRNA,
yet be highly functional. If this kind of collective regulation does exist,
it would imply a reexamination of how mutation rate can be used as a criterion for when ncRNA
is under evolutionary constraint. It should also be noted that this feature of high mutation
rate makes such massive parallelism unlikely in protein regulatory networks, which also
have clear similarities with neural networks~\cite{bray1995protein}. 

In this kind of collective mechanism, we expect that typically there will be weak influences between any two RNA molecules,
and also very many such weak interactions. This makes it difficult to reconstruct the circuit
diagram. In addition, molecules that are involved with this kind of collective
regulation could have other functions, making it hard to identify specific interactions
that support this picture. However the prevalence of ncRNA-ncRNA binding has not been the focus
of much research.
It would therefore be interesting to probe the statistics of ncRNA-ncRNA binding 
in vitro and in vivo. There is already work that has investigated the RNA-RNA interactions for
two specific cases of ncRNA~\cite{engreitz2014rna} using RNA antisense purification. It would be interesting to 
extend this to a larger class of ncRNA and determine how prevalent ncRNA-ncRNA interactions are experimentally. 
If indeed there are a large number of weak interactions, it would make the possibility discussed here
much more promising.

This work was supported by the Foundational Questions Institute \url{<http://fqxi.org>}.

\bibliography{boltz_genome}

\end{document}